\documentclass[aps,prfluids,11pt,superscriptaddress,showpacs,floatfix,amsmath,amssymb,longbibliography]{revtex4-1}
\usepackage{color,graphicx,array,tabularx,bm,url}
\newcolumntype{x}[1]{>{\centering\arraybackslash\hspace{0pt}}p{#1}}

\begin{document}

\title{Drag enhancement in a dusty Kolmogorov flow}

\author{A. Sozza}
\thanks{Corresponding author}
\email{asozza.ph@gmail.com} 
\affiliation{Istituto dei Sistemi Complessi, CNR, via dei Taurini 19, 00185 Rome, Italy 
and INFN sez. Roma2 "Tor Vergata"}

\author{M. Cencini}
\thanks{Corresponding author}
\email{massimo.cencini@cnr.it} 
\affiliation{Istituto dei Sistemi Complessi, CNR, via dei Taurini 19, 00185 Rome, Italy 
and INFN sez. Roma2 "Tor Vergata"}

\author{S. Musacchio}
\affiliation{Dipartimento di Fisica and INFN, Universit\'a di Torino, via P. Giuria 1, 10125 Torino, Italy.}

\author{G. Boffetta}
\affiliation{Dipartimento di Fisica and INFN, Universit\'a di Torino, via P. Giuria 1, 10125 Torino, Italy.}

\begin{abstract}
Particles suspended in a fluid exert feedback forces that can significantly 
impact the flow, altering the turbulent drag and velocity fluctuations. 
We study flow modulation induced by particles heavier than 
the carrier fluid in the framework of an Eulerian two-way 
coupled model, where particles are represented by a continuum 
density transported by a compressible velocity field, exchanging 
momentum with the fluid phase. We implement the model in direct 
numerical simulations of the turbulent Kolmogorov flow, a simplified 
setting allowing for studying the momentum balance and the turbulent 
drag in the absence of boundaries. We show that the amplitude of the 
mean flow and the turbulence intensity are reduced by increasing 
particle mass loading with the consequent enhancement of the friction 
coefficient. Surprisingly, turbulence suppression is stronger for 
particles of smaller inertia. We understand such a result by mapping 
the equations for dusty flow, in the limit of vanishing inertia, to a 
Newtonian flow with an effective forcing reduced by the increase in 
fluid density due to the presence of particles. We also discuss the 
negative feedback produced by turbophoresis which mitigates the 
effects of particles, especially with larger inertia, on the turbulent flow.
\end{abstract}

\date{\today}

\maketitle

\section{Introduction}
\label{sec:intro}

Dust and particulate in turbulent flows are common to many natural 
environments \cite{balmforth2001}, from aerosol in clouds formation 
\cite{shaw2003particle,celani2005}, particle-driven gravity currents 
\cite{necker2005}, sediment transport in rivers \cite{burns2015}, 
volcanic eruptions \cite{bercovici2010}, to planetesimals and 
proto-planets formation \cite{homann2016,fu2014}. They are also 
relevant to many industrial processes dealing with pipe flows and 
open channel flows \cite{lashgari2014}, as well as in fluidization 
processes \cite{bi2000}.

Dispersed particles are not only transported by the flow, but they 
exert forces (e.g. drag forces) on the fluid that, depending on the 
mass loading, can modify the flow itself. The coupled system made of 
the carrier fluid and the particles is generally referred to as 
particle-laden flow \cite{balachandar2010}. The interactions between 
the particles and the fluid can significantly alter the flow both at 
large and small scales. In particular, heavy particles can attenuate 
or enhance turbulence depending on their size with respect to the 
viscous scale \cite{gore1989effect,balachandar2010}. In general, 
smaller \cite{eaton2009two,kulick1994particle} and settling 
\cite{monchaux2017,kasbaoui2019,muramulla2020} particles lead to 
turbulence attenuation. Less clear is the effect on turbulent drag: 
experiments in channel flows did not find measurable changes on 
the mean flow \cite{eaton2009two,kulick1994particle}, while simulations 
reported drag reduction in a channel flow \cite{zhao2010} and drag 
enhancement in an unstably stratified boundary layer \cite{li2019drag}, 
moreover the effects depend sensitively on many factors including particle shape, 
size and volume/mass fraction \cite{ardekani2017,fornari2016effect}. 
At small scales, effects of particles on the carrier fluid have been observed 
in the spectral distribution of the fluid kinetic energy 
\cite{gualtieri2017turbulence,bec2017,pandey2019}.

Turbulence in multiphase flows constitutes a formidable challenge 
even in the dilute regime, where the fluid-particle interactions 
causes also the formation of strong inhomogeneities in particles' 
spatial distribution \cite{balachandar2010}. 
Fractal clustering of (one-way coupled) particles has been observed 
at small scales in chaotic flows \cite{balkovsky2001,bec2003} 
and within the inertial and dissipative range of turbulence \cite{bec2007}. 
In inhomogeneous turbulent flows large-scale clustering of particle 
occurs because of the turbophoresis, that is, the migration of the 
particles in regions of lower turbulence intensity 
\cite{Sehmel1970,Caporaloni1975,reeks1983transport,belan2014localization}. 
Due to its importance for applications, turbophoresis is usually 
studied in the presence of boundaries, such as in turbulent boundary layers 
\cite{Sehmel1970,Caporaloni1975,Brooke1992,Marchioli2002,Sardina2012}, 
pipe flows \cite{picano2009spatial} and channel flows \cite{rouson2001preferential}. 
Nevertheless, turbophoresis does not require the presence of boundaries, 
but just the spatial modulation of the turbulent intensity, and has been observed 
also in the absence of walls \cite{bec2014,delillo2016,mitra2018turbophoresis}.

In this paper, we investigate the effects of mass loading and particle 
inertia on turbulent drag and turbophoresis in bulk flows without 
material boundaries, in the regime of low volume fraction. 
To this aim we have performed numerical simulations of a two-way 
coupled fully Eulerian model, first introduced by Saffman \cite{saffman1962}, 
for a dilute suspension of inertial particles in a turbulent Kolmogorov flow. 
The Kolmogorov flow is obtained by forcing the Navier-Stokes equations 
with a sinusoidal force, and was originally proposed by Kolmogorov as a 
model to understand the transition to turbulence \cite{sivashinsky1985weak}. 
It represents a paradigm of inhomogeneous turbulent flows without boundaries, 
because the local intensity of turbulent fluctuations is spatially modulated 
by the presence of a sinusoidal mean velocity profile. 
Owing to the spatial variation of the turbulent intensity, the Kolmogorov flow 
provides an ideal setup to study the turbophoretic effect in the absence of 
boundaries \cite{delillo2016,garg2018}. 
Furthermore, the presence of a mean flow allows to define a drag (or friction) 
coefficient, as the ratio between the work made by the force 
and the kinetic energy carried by the mean flow \cite{musacchio2014}. 
In this regard, the Kolmogorov flow can be thought as a simplified channel flow 
without boundaries, and it has been exploited for numerical studies of the bulk 
processes of drag reduction in dilute polymer solutions \cite{boffetta2005}, 
drag enhancement in dilute solutions of inextensible rods 
\cite{emmanuel2017emergence} and in spatially fixed networks of rigid fibers 
\cite{olivieri2020turbulence}.

We find that particles modify the bulk properties of the flow by reducing 
the amplitude of the mean flow and the intensity of turbulent fluctuations, 
at increasing the mass loading. The reduced mean flow at fixed forcing 
amplitude implies an increase of the drag coefficient. Surprisingly, we find 
that this effect is larger for particles with smaller inertia. 
Turbulence reduction at increasing mass loading also results in a reduction 
of the turbophoretic effect, in agreement with previous findings in channel flows
\cite{grigoriadis2011reduced}.

The manuscript is organized as follows. In Sec.~\ref{sec:mod}, we 
describe the Eulerian model for a dusty fluid. In Sec.~\ref{sec:sim}, 
we detail the numerical implementation of the model and report the 
parameters used in the simulations. In Sec.~\ref{sec:res}, we present 
the main results of our study. Finally, in Sec.~\ref{sec:outro} 
we summarize the results and discuss the perspectives of our study.

\section{Eulerian model for a dusty fluid}
\label{sec:mod}

Theoretical and numerical studies of particle laden flows make use of
different models to describe the interactions between particles and
fluid \cite{balachandar2010}, based either on Eulerian-Lagrangian
approaches (see, e.g., \cite{gualtieri2017turbulence,pandey2019}) or Eulerian
two-phase models (e.g., \cite{saffman1962,bec2017}). Here we adopt an
Eulerian model with two-way coupling appropriate for suspensions with
negligible volume fraction, which was first introduced by Saffman to study the
linear stability of a dusty gas \cite{saffman1962}.

We consider a dilute mono-disperse suspension of small, heavy
particles with density $\rho_p$ and size $a$ transported in a
Newtonian fluid with density $\rho_f$ and viscosity $\mu$. 
The particle density is assumed to be much larger than fluid one, 
$\rho_p \gg \rho_f$. In real systems, the density ratio $\rho_p / \rho_f$ 
can easily reach order $10^3$ for grains or water droplets in air and
order $10$ for metallic particles in water. We assume the the 
particle size much smaller than the viscous scale of the flow, 
$a \ll \eta$, where $\eta = (\nu^3/\varepsilon)^{1/4}$ is the Kolmogorov
viscous length and $\varepsilon$ the fluid kinetic energy dissipation
rate. This assumption implies that the particle Reynolds number is
small and we further assume that the volume fraction of the 
particles $\Phi_v = N_p v_p / V$, defined in terms of the volume 
of each particle $v_p \propto a^3$ and the number of particles 
$N_p$ contained in the total volume $V$, is negligible small. 
Even for very small volume fraction, the mass loading 
$\Phi_m = \Phi_v \rho_p/\rho_f $ can be of order unity because of 
the large density ratio. As an example, for a dilute suspension of 
droplets of water in air with $\Phi_v \approx 10^{-3}$ 
one has $\Phi_m\approx 1$.

Because of the vanishing volume fraction of the particles, the 
fluid density field can be assumed to be constant and, therefore, 
the velocity field of the fluid phase ${\bm u}(\bm x,t)$ 
incompressible (${\bm \nabla}\cdot{\bm u}=0$). 
The solid phase is described by the particles' velocity field 
${\bm v}(\bm x,t)$ and the normalized number density field 
$\theta(\bm x,t) = n(\bm x,t) /(N_p/V)$, where $n(\bm x,t)$ 
is the local number of particles per unit volume. 
The normalization gives $\langle \theta \rangle = 1$. 
Here and in the following, the brackets $\langle [\cdot] \rangle$ 
denote the average over the whole volume $V$. 

For small volume fractions ($\Phi_v < 10^{-3}$) the dynamics of the
particle-laden flow can be described by a two-way coupling, which
takes into account the interactions between individual particles and
the surrounding flow, but neglects the interactions between particles
(collisions and friction) and the particle-fluid-particle interactions
(fluid streamlines compressed between particles)
\cite{elghobashi1994predicting}. In the two-way coupling regime, the
exchange of momentum between the two phases can no longer be 
neglected \cite{balachandar2010}. For small heavy particles, 
such an exchange is mainly mediated by the viscous drag force 
${\bm f}_{drag} = \gamma ({\bm v} - {\bm u})$, which is proportional 
to the velocity difference between particle and fluid velocity, 
$\gamma$ being the viscous drag coefficient.

Assuming that the interactions conserve the total momentum, 
Saffman \cite{saffman1962} derived the following 
coupled equations for the two phases:
\begin{eqnarray}
\partial_t {\bm u} + {\bm u}\cdot{\bm \nabla}{\bm u} 
&=& - \frac{{\bm \nabla}p}{\rho_f} + \nu {\bm \nabla}^2 {\bm u} 
+ {\bm f}_{ext} + \dfrac{\Phi_m}{\tau}\theta({\bm v}-{\bm u}) 
\label{eq1.5a}
\\
\partial_t {\bm v} + {\bm v}\cdot{\bm \nabla}{\bm v}\; 
&=& -\dfrac{{\bm v}-{\bm u}}{\tau} 
\label{eq1.5b}
\\
\partial_t \theta + {\bm \nabla}\!\cdot\!({\bm v}\theta) &=& 0 
\label{eq1.5c}\,,
\end{eqnarray}
where $p$ is the pressure, ${\bm f}_{ext}$ is the external force which
sustains the flow, $\nu = \mu/\rho_f$ is the kinematic viscosity, and
$\tau = m_p/\gamma$ is the particle relaxation time, defined as the
ratio between the particle mass $m_p= \rho_p v_p$ and its viscous drag
coefficient $\gamma$. In the case of a spherical particles of radius
$a$ one has $m_p = (4/3) \pi a^3 \rho_p$ and $\gamma = 6 \pi \mu a$,
which gives the Stokes time $\tau = (2/9) a^2 \rho_p/\rho_f \nu$. 
Normalizing the latter with the Kolmogorov viscous time, 
$\tau_\eta = (\nu/\varepsilon)^{1/2}$, we obtain the Stokes number 
$St=\tau/\tau_\eta$, which provides a non-dimensional measure of 
particle inertia in responding to the fluid velocity fluctuations.

It is important to remark that the validity
of the model (\ref{eq1.5a}-\ref{eq1.5c}) is limited to small Stokes
numbers $St<1$. In a Lagrangian description, nearby particles with
large $St$ may exhibit very different velocities \cite{bec2010}, 
a phenomenon known under the name of caustics formation
\cite{wilkinson2005} and sling effect \cite{falkovich2002}. 
Within the Eulerian framework, caustics would imply a multi-valued 
particle velocity field, breaking the validity of the continuum description.
The rate of caustic formation increases with $St$
\cite{wilkinson2006caustic}, therefore the Eulerian description for
the particles is valid only for sufficiently small inertia, when the
effect of caustics is negligible. A direct comparison of the model
(\ref{eq1.5a}-\ref{eq1.5c}) (for $\Phi_m=0$) with Lagrangian
simulations has shown that the Eulerian and Lagrangian approaches are
equivalent for $St<1$ \cite{boffetta2007}. Moreover, in
Eq.~(\ref{eq1.5b}) we have neglected the gravity acceleration 
${\bm g}$ on the particles to avoid additional effects induced by
sedimentation.

Noticing that in the case of spherical particles the Stokes number 
can be written as $St=(2/9)(\rho_p/\rho_f)(a/\eta)^2$, it is easy to
realize that the condition $St<1$ for the validity of the Eulerian 
description can be fulfilled only by very small particles with $a \ll \eta$. 
In order to obtain finite values for the parameters $\tau$ and
$\Phi_m$, the limit of vanishing radius $a \to 0$ can be consistently
achieved in the model (\ref{eq1.5a}-\ref{eq1.5c}) by assuming the
scaling $\rho_p / \rho_f \sim a^{-2}$ for the density ratio and 
$N_p \sim a^{-1}$ for total number of particles. 
These scalings ensure the volume fraction to vanish as $\Phi_v \sim a^2$.

We also remark that the two-way coupling used in the above model
does not preserve the total kinetic energy of the fluid and particle phases.
Defining the kinetic energy per unit volume as
$E = \rho_f (\langle |{\bm u}|^2\rangle +
\Phi_m \langle \theta|{\bm v}|^2\rangle)/2$ 
the energy balance of the model is:
\begin{equation}
\frac{dE}{dt} = \rho_f \left[ -\nu \langle ({\bm \nabla}{\bm u})^2\rangle 
- \frac{\Phi_m}{\tau} \langle \theta |{\bm v}-{\bm u}|^2\rangle 
+ \langle {\bm f}_{ext}\cdot{\bm u}\rangle \right]\;,
\label{eq:1.6}
\end{equation}
which shows that a fraction of the energy injected by the external force
is removed by the viscous drag between the particles and the fluid.

In this paper, as for the external force stirring the fluid, we consider the
Kolmogorov force ${\bm f}_{ext} = F \cos(Kz) \hat{{\bm x}}$. 
Under this forcing one has a simple laminar solutions to 
(\ref{eq1.5a}-\ref{eq1.5c}) given by $\theta=1$ and
${\bm u} = {\bm v} = U_0 \cos(Kz) \hat{{\bm x}}$ with 
$U_0=F/(\nu K^2)$.
In the absence of particles ($\Phi_m=0$), this solution becomes 
unstable to transverse
large-scale perturbations (for wavenumber smaller than $K$) when the
Reynolds number ${\rm Re}=U_0/(\nu K)$ exceeds the critical threshold
${\rm Re}_c = \sqrt{2}$ \cite{meshalkin1961}. Remarkably, even in the
turbulent regime, the Kolmogorov flow maintains a monochromatic mean
velocity profile $\overline{\bm u} = U \cos(Kz) \hat{{\bm x}}$
with an amplitude $U$ smaller than the laminar solution $U_0$
(here and in the following the over-bar
$\overline{[\cdot]}$ denotes the average over time $t$ and
over the $x$ and $y$ coordinates). 
The presence of a non-vanishing mean velocity profile 
allows us to define the turbulent drag coefficient \cite{musacchio2014}
$f=F/(KU^2)$, in analogy with channel flows.

In summary, the dimensionless parameters which control the dynamics of the model
are the mass loading $\Phi_m = \Phi_v \rho_p/\rho_f$, the Reynolds
number $Re=U/( \nu K)$, defined in terms of the amplitude $U$ of the
turbulent mean profile of the x-component of the velocity, and the
Stokes number $St=\tau/\tau_\eta$.

\section{Numerical Simulations}
\label{sec:sim}

We performed numerical simulations of Eqs.~(\ref{eq1.5a}-\ref{eq1.5c})
by means of a $2/3$ de-aliased pseudo-spectral solver with $2^{nd}$
order Runge-Kutta time marching in a triply periodic cubic domain of
side $L=2\pi$ and grid resolution $M=256$. Small scale resolution of
the fields was ensured by requiring $k_{max}\eta \geq 2.7$
($k_{max}=M/3$). We explored three values of Stokes time
$\tau=(0.10,0.34,0.58)$ and three values of mass loading
$\Phi_m=(0.0,0.4,1.0)$, which compose a dataset of $9$ configurations
in the parameters space. The simulations with $\Phi_m=0$ correspond to
the case with passive inertial particles, previously studied in
\cite{delillo2016} using a Lagrangian scheme, whose results were used
to benchmark the Eulerian model. 
We notice that the values of dimensionless parameters $Re$ and $St$
depends also on the mass loading $\Phi_m$ and are therefore determined
{\it a posteriori} in the simulations. 
The main parameters of our simulations are summarized in Table~\ref{table:param}.

In each run we let the simulations evolve to reach a stationary state,
discarding transient behaviors. The particles were initialized with a
homogeneous density field ($\theta =1$) and velocity field equal to the
fluid one (${\bf v}={\bf u}$). After the transient, we collected $360$
profiles and fields, over a temporal series of $500$ eddy turnover
time, in order to ensure statistical convergence. The statistical
uncertainties (represented by the error-bars in the figures) have been
estimated using the variations observed by halving the statistics. In
order to avoid the development of instabilities due to strong density
gradients, which are unavoidable due to particle clustering, we added
a numerical regularization to Eqs.~(\ref{eq1.5b}-\ref{eq1.5c}). 
In particular, we considered an additional viscous term 
$\nu_p \nabla^2 {\bm v}$ and diffusivity $\kappa_p \nabla^2 \theta$ 
for the particle velocity and density field, respectively. To reduce the 
number of parameters, we fixed $\nu_p=\kappa_p=\nu$.

\begin{table}[t!]
\begin{tabular}{x{1cm}x{1cm}x{1cm}x{1cm}x{1cm}x{1cm}x{2cm}x{1cm}x{2cm}x{1cm}x{1cm}x{1cm}}
\hline \hline
Run & $\tau$ & $\Phi_m$ & $U$ & $A$ & $u'_{rms}$ & $B$ & $\theta'_{rms}$ & $\varepsilon$ & $\tau_\eta$ & ${Re}$ & ${St}$ \\
\hline
A1 & $0.10$ & $0.0$ & $0.232$ & $0.020$ & $0.199$ & $6.53 \times 10^{-3}$ & $0.264$ & $9.3 \times 10^{-4}$ & $1.04$ & $232$ & $0.10$ \\
A2 & $0.10$ & $0.4$ & $0.195$ & $0.016$ & $0.164$ & $4.29 \times 10^{-3}$ & $0.185$ & $4.8 \times 10^{-4}$ & $1.44$ & $195$ & $0.07$ \\
A3 & $0.10$ & $1.0$ & $0.160$ & $0.012$ & $0.134$ & $2.97 \times 10^{-3}$ & $0.133$ & $2.7 \times 10^{-4}$ & $1.93$ & $160$ & $0.05$ \\
\hline
B1 & $0.34$ & $0.0$ & $0.233$ & $0.047$ & $0.199$ & $6.62 \times 10^{-3}$ & $0.634$ & $9.3 \times 10^{-4}$ & $1.04$ & $233$ & $0.33$ \\
B2 & $0.34$ & $0.4$ & $0.197$ & $0.039$ & $0.160$ & $4.34 \times 10^{-3}$ & $0.444$ & $4.2 \times 10^{-4}$ & $1.54$ & $197$ & $0.22$ \\
B3 & $0.34$ & $1.0$ & $0.169$ & $0.030$ & $0.131$ & $3.02 \times 10^{-3}$ & $0.324$ & $2.4 \times 10^{-4}$ & $2.06$ & $169$ & $0.17$ \\
\hline
C1 & $0.58$ & $0.0$ & $0.233$ & $0.061$ & $0.199$ & $6.68 \times 10^{-3}$ & $0.922$ & $9.3 \times 10^{-4}$ & $1.04$ & $233$ & $0.56$ \\
C2 & $0.58$ & $0.4$ & $0.200$ & $0.048$ & $0.158$ & $4.59 \times 10^{-3}$ & $0.634$ & $4.0 \times 10^{-4}$ & $1.058$ & $200$ & $0.37$ \\
C3 & $0.58$ & $1.0$ & $0.174$ & $0.038$ & $0.129$ & $3.18 \times 10^{-3}$ & $0.458$ & $2.2 \times 10^{-4}$ & $2.12$ & $174$ & $0.27$ \\
\hline \hline
\end{tabular}
\caption{Simulation parameters: Run index, Stokes time $\tau$, 
mass loading $\Phi_m$, amplitude of the mean flow $U$, 
amplitude of the modulation of the particle density profile $A$, 
Root mean square (RMS) velocity fluctuations $u'_{rms}$, 
amplitude of the modulation of the profile of square velocity fluctuations $B$, 
RMS particle density fluctuations $\theta'_{rms}$, 
energy dissipation rate $\varepsilon = \nu \langle ({\bf \nabla}{\bf u})^2 \rangle$, 
Kolmogorov time $\tau_\eta = (\nu/\varepsilon)^{1/2}$,
Reynolds number $Re=U/(K\nu)$, Stokes number $St=\tau/\tau_\eta$.
In all runs we used resolution $M=256$, kinematic viscosity $\nu = 10^{-3}$,
forcing amplitude $F=8 \times 10^{-3}$, forcing wave-number $K=1$.}
\label{table:param}
\end{table}

Finally, we observe that in principle the pseudo-spectral scheme does 
not preserve the positivity of the density field. 
Indeed, in low density regions steep gradients and fluctuations of 
density may occasionally generate events with negative density.
Nonetheless we have checked that, even in the worst 
cases corresponding to small $\Phi_m$s and large $\tau$s,
the fraction of points with negative density does not exceeds $1-2\%$.

\section{Results}
\label{sec:res}

\begin{figure}[h!]
\centering
\includegraphics[width=\textwidth]{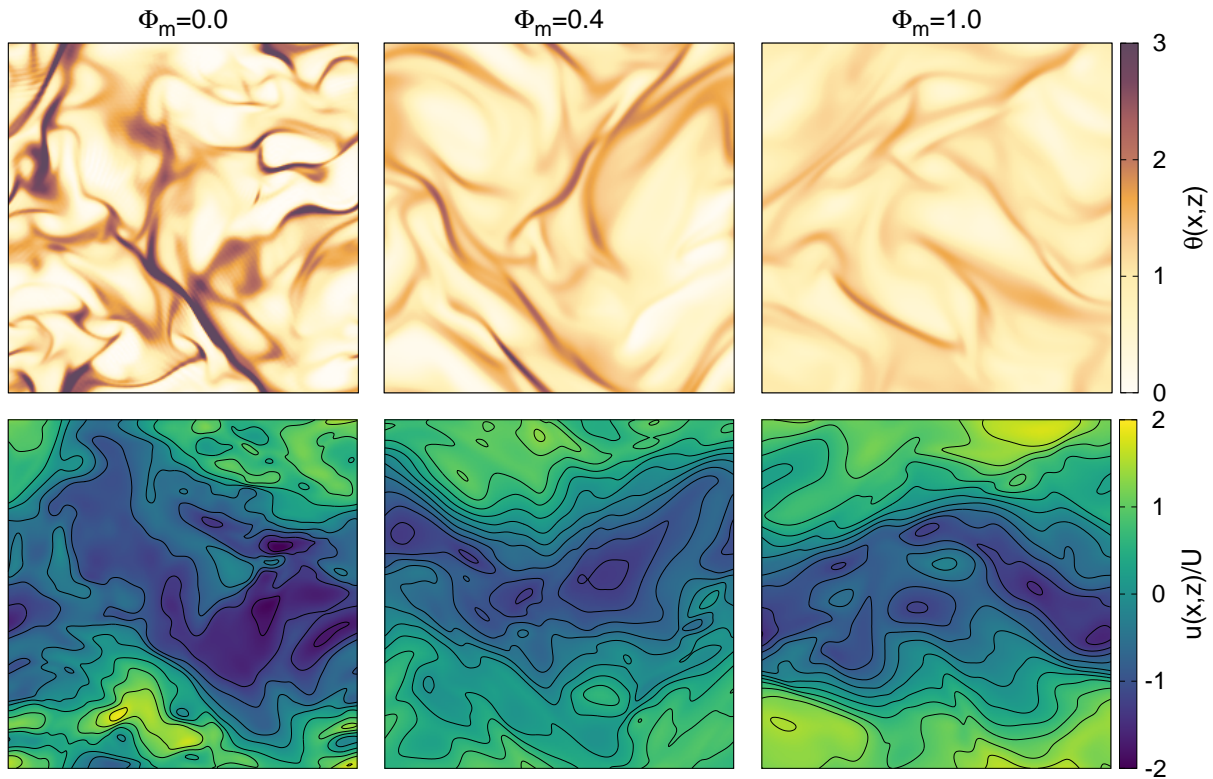}
\caption{(Color online) Visualization of two-dimensional sections in
 the plane $(x,z)$ (at fixed $y=L/2$) of the particle density field
 $\theta$ (top), and longitudinal velocity field $u_x$ (bottom)
 normalized with the amplitude of the mean flow $U$. Simulations
 refer to $\tau=0.34$ and $\Phi_m$ as labeled.}
\label{fig:sect}
\end{figure}

We start discussing the numerical results by showing, in
Fig.~\ref{fig:sect}, the two-dimensional sections of the particle
density field $\theta(x,z)$, and the longitudinal velocity field
$u_x(x,z)$ for a given Stokes time $\tau=0.34$ and different values of
the mass loading $\Phi_m$. We notice that the density field is 
organized in elongated filaments, which are gradually smoothed for
increasing mass loading. Moreover, they seem to be disposed parallel
to the isolines of the longitudinal velocity $u_x$, and correlated
with regions of strong gradients of the velocity field, where the
space between isolines is narrowed. Also the fluctuations of the
longitudinal velocity $u_x$ appear to be suppressed with respect to
the intensity of the mean flow $U$ at increasing mass loading.
Already at a qualitative level, these observations provide a first
indication that turbulence in the fluid phase is reduced by the
back-reaction of the solid phase.

Due to the symmetries of the forcing, which depends on the transverse direction $z$ only, 
we can define a mean velocity profile $\overline{\bm u}(z)$
by averaging the velocity field ${\bm u}(x,y,z,t)$
over the coordinates $x$, $y$ and time $t$.
Alike the forcing, also the mean velocity profile
has non-zero component only in the $x$-direction:
$\overline{\bm u}(z) = (\overline{u}_x(z),0,0)$.
Furthermore, we decompose the velocity field as the sum of the
mean velocity profile and the velocity fluctuations: 
${\bm u} = \overline{\bm u}+{\bm u}'$.

\begin{figure}[h!]
\centering
\includegraphics[width=\textwidth]{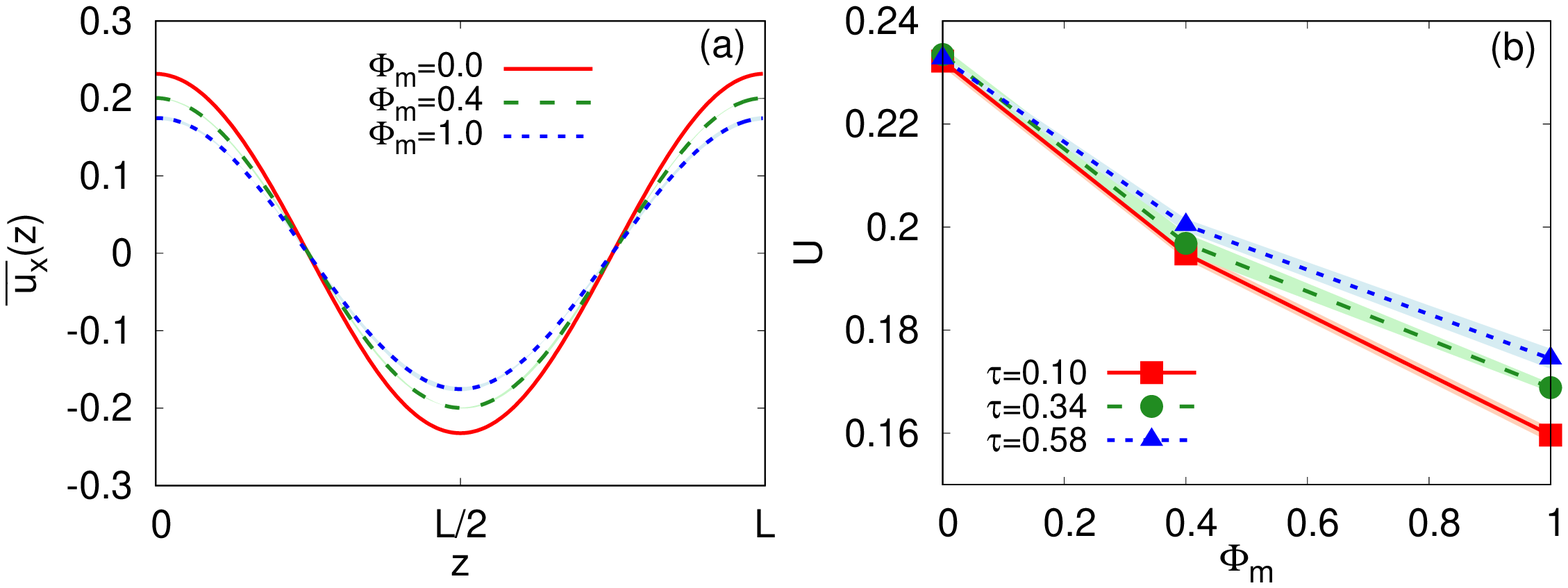}
\caption{(Color online) Averaged profiles and amplitudes of the 
longitudinal fluid velocity. (a) Mean velocity profile 
$\overline{u_x}(z)$ for different mass loading 
$\Phi_m=(0.0,0.4,1.0)$ and fixed $\tau=0.58$. (b) Amplitude of the 
mean flow $U$ as a function of $\Phi_m$ for different Stokes time as in label.}
\label{fig:prof_u}
\end{figure}

In Fig.~\ref{fig:prof_u}a we show the average profiles of the
longitudinal velocity $\overline{u}_x(z)$ for $\tau=0.58$. 
Similarly to the case of pure fluid ($\Phi_m=0$) \cite{musacchio2014},
we find that the profile of the mean flow is, with a good approximation,
monochromatic 
\footnote{For $\Phi_m >0$ deviations from the monochromatic form 
can appear because of the exchange term $\overline{\theta(v-u)}$. 
Expanding the density field as $\theta = 1+A\cos(2Kz)+o(A)$ in terms 
of the small amplitude $A$ of the density profile, the exchange term 
is, at leading order, proportional to the velocity difference 
$\overline{(v-u)}$, which gives a monochromatic velocity profile. 
Deviations from the monochromatic velocity profile appears only at 
higher order in $A$.}:
\begin{equation}
\overline{u}_x(z)=U\cos(Kz)\,. 
\label{eq:meanux}
\end{equation}
As shown in Fig.~\ref{fig:prof_u}b, the amplitude $U$ of the mean
velocity profile decreases at increasing the mass loading $\Phi_m$ 
(of about $30\%$ in the case with $\Phi_m=1$ and $\tau=0.10$). 
Even though the dependence of $U$ on $\tau$ at fixed $\Phi_m$ 
appears to be milder, Fig.~\ref{fig:prof_u}b shows that the mean flow 
is reduced more at smaller $\tau$. In other words, particles with small inertia
seem to affect more the mean flow, which is somehow counterintuitive.

\begin{figure}[t!]
\centering
\includegraphics[width=0.7\textwidth]{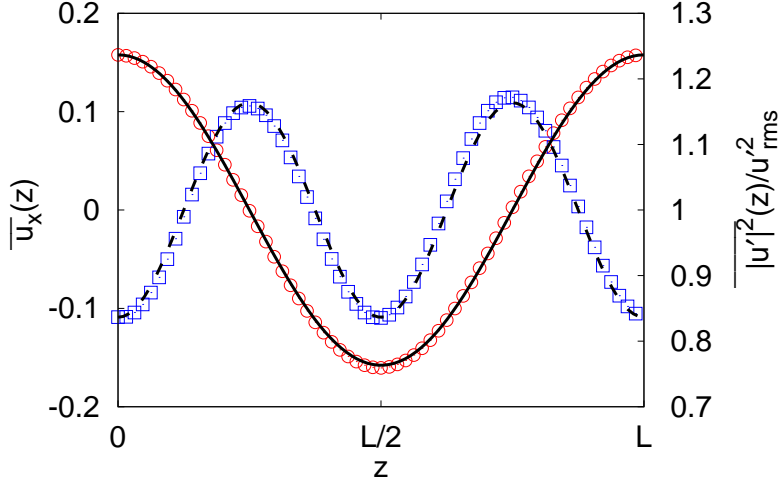}
\caption{(Color online)
Profiles of the mean longitudinal flow $\overline{u_x}(z))$ (red circles and solid line)
and the square velocity fluctuations $\overline{|{\bm u}'|^2}(z)$ (blue squares and dashed line),
for a simulation of the particle-laden flow with $\tau=0.1$, $\Phi_m=1.0$ (symbols)
and a simulation of a pure fluid with rescaled forcing amplitude $F' = F/(1+\Phi_m)$
(black lines, data from \cite{musacchio2014}).}
\label{fig:comp}
\end{figure}

The effects of the particles at small $St$ can be explained as follows. 
When the dust is sufficiently fine, i.e. $\tau \ll \tau_\eta$, 
particles follow the fluid velocity almost like tracers.
From Eq.~(\ref{eq1.5b}), at the first order in $\tau$ one can write
${\bm v} = {\bm u} - \tau D_t {\bm u} + o(\tau)$ \cite{balkovsky2001},
where $D_t = \partial_t + {\bm u} \cdot {\bm \nabla}$ represents the
material derivative. At zero order in $\tau$, the particle velocity
field remains incompressible and therefore the particle are
homogeneously distributed: $\theta = 1 + O(\tau)$. Substituting the
expansions for ${\bm v}$ and $\theta$ in Eq.~(\ref{eq1.5a}), the
equation for the fluid velocity at leading order becomes
\begin{equation}
(1+\Phi_m) D_t {\bm u} = 
- {\bm \nabla} p + \nu \nabla^2 {\bm u} + {\bm f}_{ext}\,.
\label{eq:saffman}
\end{equation}
In other terms the fluid density is increased by the presence of particles.
At low Reynolds numbers such as in the case of linear stability problems, 
as previously discussed by Saffman \cite{saffman1962}, the particle-laden 
flow is equivalent to a Newtonian fluid with a rescaled viscosity 
$\nu' = \nu/(1+\Phi_m)$ and therefore particles have a destabilizing effect. 
Conversely, at high Reynolds numbers, the viscous term is negligible 
in the momentum budget and the factor $(1+\Phi_m)$ rescales the amplitude 
of the forcing ${\bm f}'_{ext} = {\bm f}_{ext}/(1+\Phi_m)$. According to this 
argument, one expects that at small $St$ and large $Re$ the main effect 
of the particles is to cause a reduction of the external forcing and consequently 
of the mean flow intensity, therefore increasing the turbulent drag. 
We have tested this prediction by comparing the simulation of the particle-laden 
flow with $\tau=0.1$ and $\Phi_m=1.0$, with a simulation of a pure fluid 
(i.e., without particles) and rescaled forcing amplitude: $F' = F/(1+\Phi_m)$. 
As shown in Fig.~\ref{fig:comp}, the profiles of the mean flow
$\overline{u}_x(z)$ and of the velocity fluctuations $\overline{|{\bm u}'|^2}(z)$ 
obtained in the two cases coincide.
  
\begin{figure}[b!]
\centering
\includegraphics[width=\textwidth]{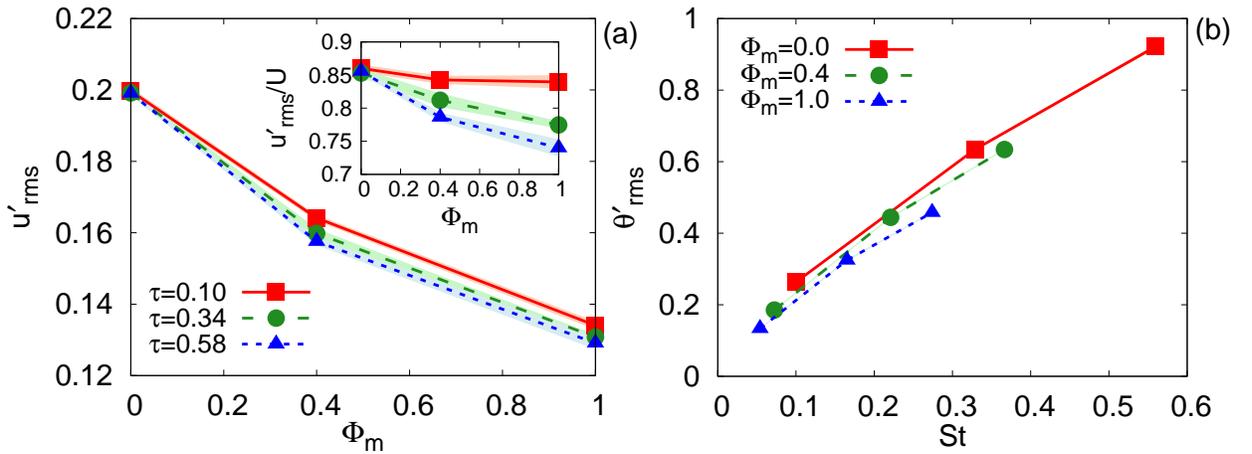}
\caption{(Color online) 
(a) RMS fluid velocity fluctuations $u'_{rms}$ as a function of $\Phi_m$.
In the inset velocity fluctuations are normalized with the mean flow. 
(b) RMS particle density fluctuations $\theta'_{rms}$ as a function of $St$.}
\label{fig:fluct}
\end{figure}

Particles impact not only on the mean flow, but also on the turbulent
fluctuations ${\bm u}' = {\bm u} - \overline{\bm u}$. 
At increasing mass loading $\Phi_m$, we observe a reduction of the 
root mean square (RMS) fluid velocity fluctuations 
$u'_{rms} = \langle |{\bm u}'|^2 \rangle^{1/2}$ (see Fig.~\ref{fig:fluct}a). 
Actually, fluctuations are suppressed more than the mean flow, 
as shown by the ratio $u'_{rms}/U$ (inset of Fig.~\ref{fig:fluct}a). 
At fixed $\Phi_m$, the dependence of $u'_{rms}$ on $\tau$ is weak 
(as for $U$) and it is opposite to what observed for $U$: 
particles with smaller $\tau$ cause a smaller reduction of $u'_{rms}$. 
In the Kolmogorov flow, the intensity of turbulent fluctuations 
is not homogeneous. Turbulence is more intense in the regions 
where the shear of the mean flow is maximum, while it is weaker 
around the maxima of the mean flow~\cite{musacchio2014,delillo2016}. 
Therefore, the profile of square velocity fluctuations displays a 
monochromatic spatial modulation: 
$\overline{|{\bm u}'|^2}(z) = ({u'_{rms}})^2 - B \cos(2Kz)$. 
As discussed in \cite{delillo2016} 
(in the case of vanishing mass loading, $\Phi_m=0$) 
the amplitude $B$ of the spatial modulation of turbulence intensity 
is directly related to the turbophoresis. The values of $B$ measured 
in our simulations are reported in Table~\ref{table:param}. 
Alike $u'_{rms}$, we find that also $B$ is strongly reduced at 
increasing $\Phi_m$ while it weakly depends on $\tau$.

\begin{figure}[t!]
\centering
\includegraphics[width=\textwidth]{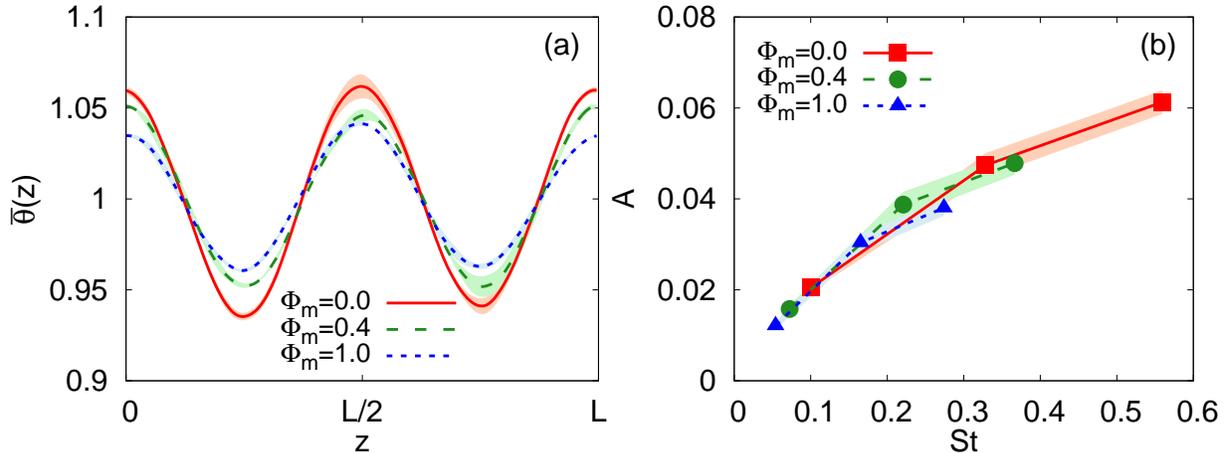}
\caption{(Color online) 
(a) Mean particle density profile $\overline{\theta}(z)$ for different values 
of mass loading $\Phi_m=(0.0,0.4,1.0)$ and fixed Stokes time $\tau=0.58$. 
(b) Amplitude $A$ of the spatial modulation of the density profile 
$\overline{\theta}(z)= 1+ A\cos(2Kz)$, as a function of the Stokes number 
$St$ for different values of $\Phi_m$.}
\label{fig:prof_rho}
\end{figure}

The turbulence attenuation caused by the mass loading reflects 
into a reduction of the turbophoretic effect.
In Fig.~\ref{fig:prof_rho}a we show that the mean particle density profile
displays a monochromatic modulation $\overline{\theta}(z)= 1+ A\cos(2Kz)$.
Note that the wavelength of the modulation of density
is equal to that of the turbulent intensity
and it is half that of the mean flow. 
For $\Phi_m=0$ the profile obtained is in agreement with the results of 
the Lagrangian simulations reported in Ref.~\cite{delillo2016}. 
The amplitude $A$ of the spatial modulation of the mean density profile
provides a quantitative measure of the turbophoretic effect.
The values of $A$ are reported in Table~\ref{table:param}
and shown in Fig.~\ref{fig:prof_rho}b.
We find that $A$ reduces at increasing the mass loading $\Phi_m$.
This effect is directly connected with the reduction of the amplitude $B$
of the variations of the turbulent diffusivity at increasing $\Phi_m$. 
Furthermore, the amplitude $A$ increases as a function of ${\rm St}$
collapsing on a master curve for all the values of $\Phi_m$.
These results shows that the coupling between the particles and the fluid
causes a reduction of the turbophoresis in the Kolmogorov flow, 
in agreement with what observed in channel flows \cite{grigoriadis2011reduced}.

In the Kolmogorov flow, the turbophoretic effect can be observed only
by long time averages of the density profiles, but it is not directly
visible in the instantaneous density fields. 
As shown in Figure~\ref{fig:sect}, the latter are characterized by 
filaments of clustered particles. Clustering intensity can be quantified 
by decomposing the particle density field as 
$\theta = \overline{\theta} + \theta' = 1+A\cos(2Kz) + \theta'$. 
The values of the RMS density fluctuations $\theta'_{rms}$ are shown 
in Fig~\ref{fig:fluct}b. Similarly to what observed for the amplitude $A$ 
of the mean density profile, we find that $\theta'_{rms}$ reduces at 
increasing mass loading $\Phi_m$. Again, this is due to the reduction 
of turbulence at increasing $\Phi_m$, which results in larger values for 
the Kolmogorov times $\tau_{\eta}$ and hence reduces the particles 
Stokes number $St =\tau/\tau_{\eta}$. 
Particle clustering is therefore suppressed by the mass loading.

The effects of the solid phase on the fluid can be further quantified 
by inspecting the equation for the local balance of fluid momentum. 
By averaging (\ref{eq1.5a}) over $x,y$ and $t$, we obtain the equation 
\begin{equation}
\partial_z \overline{u_x u_z} - \nu \partial_{zz} \overline{u_x} 
- F \cos(K z) - \dfrac{\Phi_m}{\tau} \overline{\theta (v_x - u_x)} = 0\,,
\label{eq4.1a}
\end{equation}
for the mean profiles of the turbulent Reynolds stress ($\overline{u_x u_z}$), 
of the viscous stress ($\nu \partial_{z} \overline{u_x}$) of the forcing ($F\cos(K z)$) 
and of the momentum exchange with the solid phase 
($\dfrac{\Phi_m}{\tau} \overline{\theta (v_x - u_x)}$). 
Because of the monochromatic forcing, we can assume at first approximation 
a monochromatic profile for the terms in Eq~(\ref{eq4.1a}), i.e. besides 
(\ref{eq:meanux}) we assume
\begin{equation}
\overline{u_x u_z} = S\sin(K z), \qquad
\overline{\theta(v_x-u_x)} = - X \cos(K z)
\label{eq4.2}
\end{equation}
where $S$ is the amplitude of the Reynolds stress and $X$ is the
amplitude of the momentum exchange. 
Following Ref.~\cite{musacchio2014}, inserting Eq.~(\ref{eq:meanux}) 
and (\ref{eq4.2}) in the momentum equation (\ref{eq4.1a}), yields the
following algebraic relation for the amplitudes
\begin{equation}
- S K -\nu K^2 U + F - \dfrac{\Phi_m}{\tau} X = 0\,.
\label{eq:mom}
\end{equation}

\begin{figure}[t!]
\centering
\includegraphics[width=\textwidth]{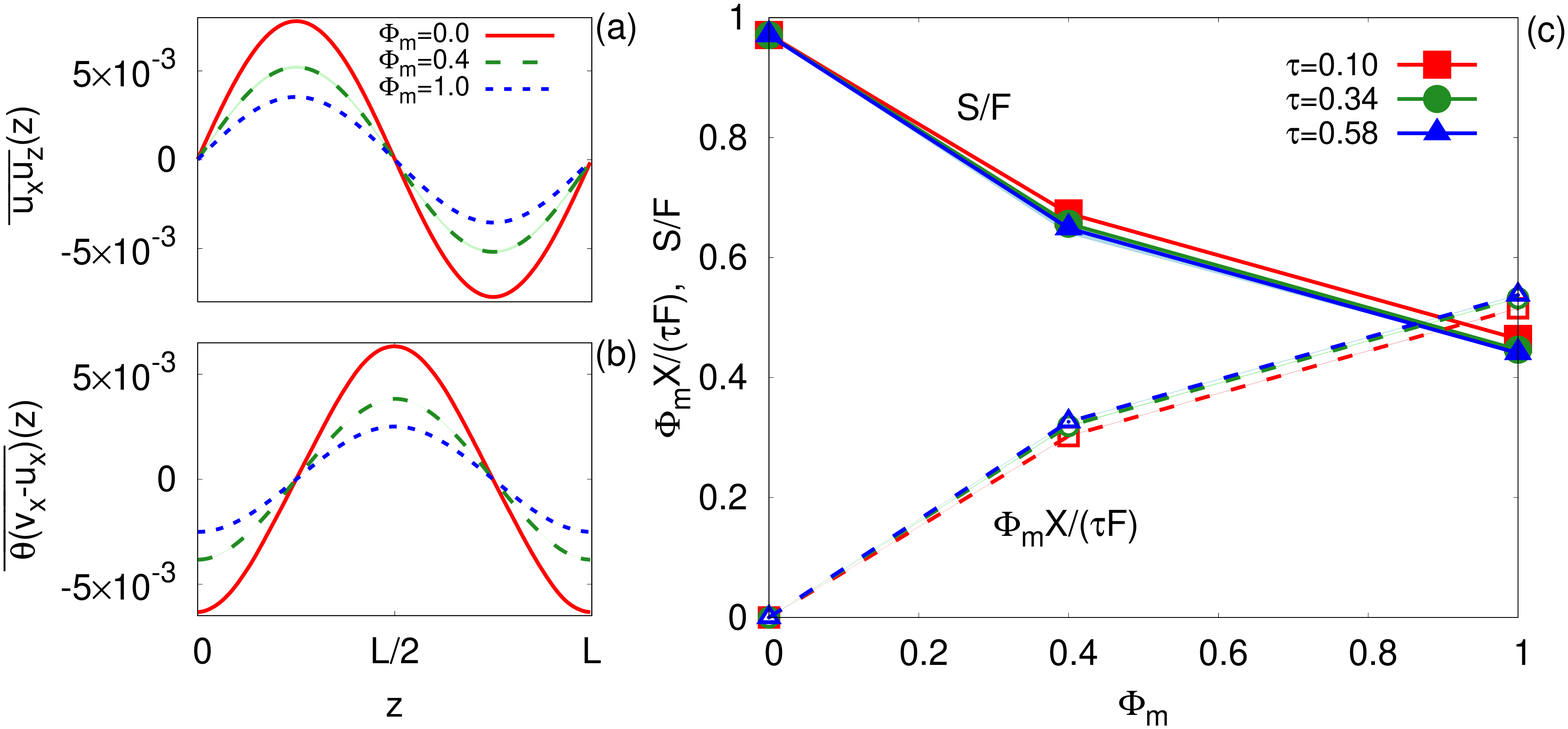}
\caption{(Color online) Momentum budget. 
Mean profile (a) of the Reynolds stress term and 
(b) of the momentum exchange term for 
different mass loadings as in label, with $\tau=0.58$. 
Panel (c) shows the momentum budget for the amplitudes 
divided into Reynolds stress term (filled symbols)
and exchange term (empty symbols),
as a function of the mass loading $\Phi_m$
for different values of Stokes times as in label.}
\label{fig:mom}
\end{figure}

In Fig.~\ref{fig:mom}a-b, we show the profiles of the Reynolds stress 
and momentum exchange for different values of $\Phi_m$ and $\tau=0.58$. 
They are very well approximated by the monochromatic functional form (\ref{eq4.2}). 
The amplitudes of the Reynolds stress and exchange terms, 
normalized with the forcing amplitude $F$, are shown in Fig.~\ref{fig:mom}c. 
The amplitude of the viscous term $\nu K^2 U/F$ (not shown) remains 
small with respect to the other terms (its effect on the total budget is only about $2-3\%$). 
For vanishing mass loading ($\Phi_m=0$), the exchange term is zero 
and the Reynolds stress contribution is maximum, while increasing 
$\Phi_m$ the two terms becomes of the same order. 
For even larger mass loading ($\Phi_m \gtrsim 1$) the coupling term dominates 
over the Reynolds stress term. Notice that the dependence on $\tau$ is very 
weak, this is consistent with the observation that, since at leading order 
${\bm v}-{\bm u} \approx - \tau D_t {\bm u}$, the amplitude of exchange term 
$X$ is order $\tau$, meaning that $\Phi_m X/\tau F$ depends upon $\tau$ only at higher orders.

\begin{figure}[t!]
\centering \includegraphics[width=\textwidth]{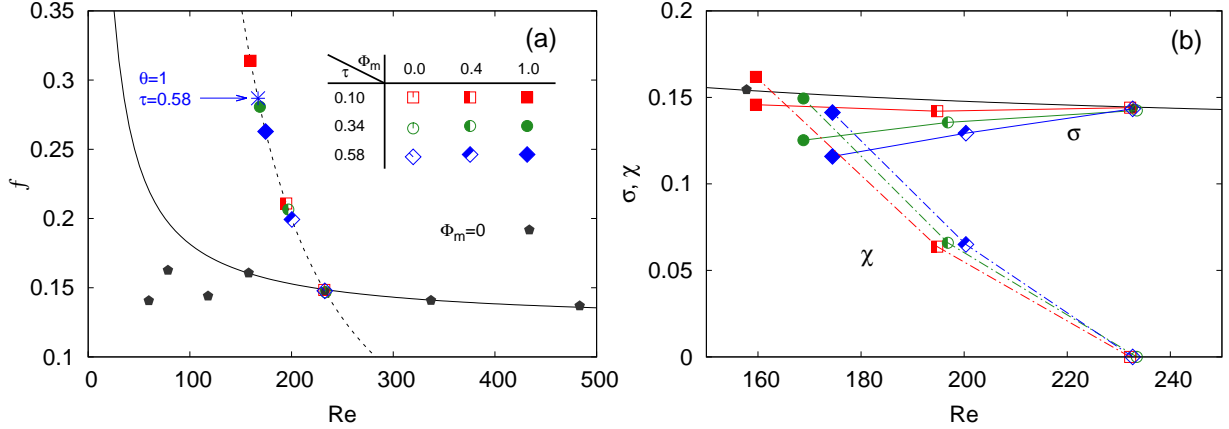}
\caption{(Color online) Friction and stress coefficients. 
(a) Friction factor $f$ (filled symbols) as a function of $Re$ for all the
parameter configurations ($\Phi_m$,$\tau$) as in legend. The black
pentagons are the values of $f$ in the absence of particles ($\Phi_m=0$)
and the black continuous line is $f=f_0+b/Re$
with $f_0=0.124$ and $b=5.75$ \cite{musacchio2014}.
Dashed line represents the curve $f=F/(\nu^2 K^3 Re^2)$.
Blue asterisk corresponds to the simulation with imposed uniform 
density ($\theta=1$) at $\Phi_m=1$ and $\tau=0.58$ (same 
parameters of run C3).
(b) Stress coefficient $\sigma$ (solid curves) and
exchange coefficient $\chi$ (dotted curves).}
\label{fig:drag}
\end{figure}

The dimensionless version of the momentum budget is obtained 
by dividing all the terms of Eq.~(\ref{eq:mom}) by $KU^2$ and 
defining the friction coefficient $f=F/(K U^2)$~\cite{musacchio2014}, 
(quantifying the ratio between the work done by the force 
and the kinetic energy of the mean flow) 
the Reynolds stress coefficient $\sigma=S/U^2$, 
and the exchange coefficient $\chi = \Phi_m X/(\tau K U^2)$: 
\begin{equation}
f = \dfrac{1}{Re} + \sigma + \chi \,.
\label{eq:fac}
\end{equation}
In Fig.~\ref{fig:drag}a, we show the friction factor $f$ as a function
of the Reynolds number. In the absence of particles ($\Phi_m=0$) 
an asymptotic constant value for the friction coefficient is reached 
for large enough Reynolds numbers as $f=f_0 + b/Re$ 
(with $f_0=0.124$ and $b=5.75$)\cite{musacchio2014}. 
Figure~\ref{fig:drag}a shows that the presence of particles increases 
the friction coefficient, by reducing the mean velocity $U$. We remark 
that, since both $f$ and $Re$ depend solely on $U$ and do not depend 
explicitly on the particle parameters $\tau$ and $\Phi_m$, all the values 
of $f$ obtained in the simulations at fixed $F$ and $\nu$ lie on the curve 
$f=F/(\nu^2 K^3 Re^2)$. Not surprisingly, the effect is stronger for larger
values of $\Phi_m$ (vanishing in the passive limit $\Phi_m=0$), 
while we find that the largest friction is obtained with smaller 
Stokes times, in particular for large $\Phi_m$. 
As discussed above (\textit{Cfr} Eq.~(\ref{eq:saffman})), this behavior 
is a consequence of the reduced effective forcing in the limit of 
vanishing inertia $St \to 0$. In this limit the velocity field ${\bm u}$ 
is equal to that of a pure fluid (without particles) which satisfies 
the Navier-Stokes equation with rescaled forcing 
${\bm f}'_{ext} = {\bm f}_{ext} / (1+\Phi_m)$ 
and viscosity $\nu' = \nu / (1+\Phi_m)$
\footnote{
 At high $Re$, the rescaling of the viscosity
 is negligible in the momentum balance,
 as confirmed by the data in Figure~\ref{fig:comp}}.
The friction factor of the dusty Kolmogorov flow is therefore: 
$f=F/(KU^2) = (1+\Phi_m) F'/(KU^2) = (1+\Phi_m) f'$, 
where $f'= F'/(KU^2)$ is the friction factor
of the pure fluid with rescaled Reynolds number 
$Re' = U/(\nu' K) = Re (1+\Phi_m) $.
For $Re \gg 1$ the friction factor $f'$ follows
the asymptotic behavior $f' = f_0 + b/Re'$. 
This leads to an expression for the friction factor of the particle-laden flow
at large $Re$ and small $St$: 
\begin{equation}
f = (1+\Phi_m) f_0 + \frac{b}{Re}\,.  
\label{eq:fric_pred}
\end{equation}
Equating the above relation with $f = F/(\nu^2 K^3 Re^2)$ we get
a prediction for $Re$ (valid for $Re \gg 1$ and $St \ll 1$) 
in terms of the parameters $F,K,\nu,\Phi_m$: 
\begin{equation}
Re = \frac{b}{2 f_0(1+\Phi_m)} 
\left[ \sqrt{1+4\frac{f_0 (1+\Phi_m) F}{b^2 \nu^2 K^3}} -1 \right]\,.
\label{eq:re_pred}
\end{equation}
The values of $Re$ obtained in our simulations with the smallest inertia ($\tau=0.10$)
are in agreement (within $5\%$) with the prediction~(\ref{eq:re_pred}). 

We now consider the behavior of the stress coefficient $\sigma$.
We remark that in absence of particles ($\Phi_m=0$),
$\sigma$ follows the expression $\sigma = f_0 + (b-1)/Re$ \cite{musacchio2014},
inherited from the Re-dependence of the friction factor $f$.
Increasing the mass loading $\Phi_m > 0$, 
$\sigma$ attains values not too far from the case $\Phi_m=0$,
but slightly shifted below.
By considering points at constant $\tau$,
they appear to be disposed in lines that stray from the
point at $\Phi_m=0$ with different slopes.
Increasing $\tau$, the lines gradually deviates from the curve at $\Phi_m=0$. 
Increasing the mass loading, $\sigma$ decreases while $\chi$ grows, 
similarly to what observed for the momentum budget in Fig~\ref{fig:mom}c.
Although, the momentum balance indicates a drastic reduction of the 
Reynolds stress $S$, 
the stress coefficient $\sigma$ shows a much weaker dependence on $\Phi_m$, 
with only moderate variations $20\%$ at most, 
with respect to the friction coefficient $f$, which is increased of about 
$110\%$.

Finally, we discuss the role of turbophoresis on the friction coefficient.
Since turbophoresis reduces the concentration of particles in the 
regions of higher turbulence intensities, it is expected to reduce, 
by a negative feedback, the effect of particles on the turbulent flows.
In order to address this point, we performed additional simulations 
of equations (\ref{eq1.5a}-\ref{eq1.5b}) in which the particle density field
is artificially imposed to be homogeneous ($\theta\equiv1$), thus 
switching off any turbophoretic effect.
The result of these simulations is shown in Fig.~\ref{fig:drag}a
for the largest Stokes time ($\tau=0.58$) and mass loading $\Phi_m=1$.
It is evident that, at given $\Phi_m$ and $\tau$, the simulation with 
imposed uniform concentration produces a larger effect (larger friction coefficient)
with respect to the fully coupled model, since it suppress the negative feedback 
produced by turbophoresis. This is in agreement with the observation that, 
at fixed $\Phi_m$, particles with larger $\tau$, displaying a larger turbophoretic effect, 
cause a weaker increase of the drag coefficient than the particles with smaller $\tau$. 

\section{Summary and perspectives}
\label{sec:outro}

In this work we have presented the results of numerical simulations of
a fully Eulerian model for a two-way coupled particle-laden turbulent
Kolmogorov flow at varying the inertia and mass loading of the
dispersed particle phase. The peculiarity of the Kolmorogov flow is
that, while it has no material boundaries, it is characterized by a
well defined mean velocity profile as well as persistent regions of
low and high turbulent intensity. These features are here exploited
to study the active role of the particles in the phenomena of drag
enhancement and turbophoresis occurring in bulk flow.

We have shown that, at increasing mass loading, the Stokes drag
exerted by particles on the fluid phase induces a reduction of both
the mean flow and the turbulent fluctuations. As a consequence, the
presence of suspended particles reduces the Reynolds number and
increases the friction coefficient, defined as the ratio between the
work of the external force and the kinetic energy of the mean flow.
Noteworthy, we have found that the drag enhancement is higher in the
case of particles with smaller inertia which, at a first glance,
appears counterintuitive because for vanishing inertia particles are
expected to recover the dynamics of fluid elements. While the latter
expectation is true, one must consider that the particles are heavier 
than the fluid. As a result, the fluid and the particles, in the limit
of vanishing inertia, basically form a denser fluid. Using this simple idea, 
originally due to Saffman \cite{saffman1962}, 
we could explain the apparently counterintuitive dependence on the
Stokes number in terms of an effective rescaling of the forcing
amplitude caused by the increase in fluid density. 
The suppression of turbulent intensity at increasing
mass loading causes a reduction of the turbophoresis, quantified by
the amplitude of the spatial modulation in the mean particle density
profile. As expected, this effect is more pronounced for particles
with large inertia. Furthermore, because of their preferential
migration toward regions of weaker turbulent intensity, particles with
large inertia are less efficient in exerting their drag on the fluid
and, therefore, they cause a weaker drag enhancement with respect to
particles with smaller inertia at equal mass loading.

It is worth to compare the effects of the particle phase in the
Kolmogorov flow with those observed in channel flows. The reduction
of the turbophoresis at increasing mass loading and turbulent
attenuation are observed both in the Kolmogorov and channel
flows~\cite{grigoriadis2011reduced,eaton2009two}. Drag enhancement
observed in the Kolmogorov flow seems to be at odds with the
observation of Ref.~\cite{zhao2010} that reported drag reduction in
channel flow simulations, however other works did not found
significant variations of the mean flow
\cite{kulick1994particle,eaton2009two}. In general, in wall bounded
flows the effects of the particles in the boundary layers might be
sensitive to details and more important than those occurring in the
bulk flow, in this respect the Kolmogorov flow provides a useful
numerical setup to investigate the latter.

Concerning the relative importance of the mass loading and inertia,
based on our numerical simulations of the particle-laden Kolmogorov
flow, we found that while the inertia plays a major role in the
particles' dynamics, it has a weaker influence on the properties of
the flow, which are more critically dependent on the mass loading. We
observe, however, that any change in the mass loading $\Phi_m$ results
also in a change of the Stokes number $St$. Indeed, an increase in
the mass loading can be achieved by (i) increasing the material
density of the particle $\rho_p$, (ii) increasing their size $a$,
(iii) increasing the number of particles $N_p$. The cases (i) and
(ii) directly imply an increase of the particle response time $\tau$,
and therefore of the Stokes number. In the case (iii) $\tau$ remains
unchanged, but the viscous time $\tau_\eta$ is affected by the
reduction of turbulent fluctuations, producing again a change of $St$.

A variety of open questions and issues here can be addressed 
using the present model. First of all, remaining within the settings 
of the Kolmogorov flow, it would be interesting to study the effect of 
particles on the stability properties at the transition from the 
laminar to the turbulent regime, where the role of particle inertia 
can be important. It would also be interesting to exploit the 
Eulerian model here discussed for studying modifications of turbulence 
at small scales extending the preliminary study of Ref.~\cite{bec2017} 
in two-dimensional turbulence and comparing with the results obtained 
with Eulerian-Lagrangian models \cite{gualtieri2017turbulence,pandey2019}. 
Moreover, the model can be easily modified to include gravity allowing 
to study sediment-laden flows \cite{burns2015} 
or particle-induced Rayleigh-Taylor instability \cite{chou2016}.

\section*{Acknowledgments}
We acknowledge HPC CINECA for computing resources (INFN-CINECA grant
no. INFN19-fldturb). GB and SM acknowledges support from the
Departments of Excellence grant (MIUR). AS acknowledges support from
grant MODSS (Monitoring of space debris based on intercontinental
stereoscopic detection) ID 85-2017-14966, research project funded by
\emph{Lazio Innova/Regione Lazio} according to Italian law L.R. 13/08.

\bibliography{biblio}

\end{document}